# Cold atom guidance in a capillary using blue-detuned, hollow optical modes

**Joseph A. Pechkis and Fredrik K. Fatemi[*]**

*Optical Sciences Division, Naval Research Laboratory, Washington DC, 20375, USA*
[*]*coldatoms@nrl.navy.mil*

**Abstract:** We demonstrate guiding of cold $^{85}$Rb atoms through a 100-micron-diameter hollow core dielectric waveguide using cylindrical hollow modes. We have transported atoms using blue-detuned light in the 1$^{st}$ order, azimuthally-polarized $TE_{01}$ hollow mode, and the 2$^{nd}$ order hollow modes ($HE_{31}$, $EH_{11}$, and $HE_{12}$), and compared these results with guidance in the red-detuned, fundamental $HE_{11}$ mode. The blue-detuned hollow modes confine atoms to low intensity along the capillary axis, far from the walls. We determine scattering rates in the guides by directly measuring the effect of recoil on the atoms. We observe higher atom numbers guided using red-detuned light in the $HE_{11}$ mode, but a 10-fold reduction in scattering rate using the 2$^{nd}$ order modes, which have an $r^4$ radial intensity profile to lowest order. We show that the red-detuned guides can be used to load atoms into the blue-detuned modes when both high atom number and low perturbation are desired.

## 1. Introduction

Atom guides using hollow optical fibers (HOFs) have continued to be of interest for potential use in nonlinear optics and optical switches [1,2], atom transport [3], and atom interferometry [4]. Guiding is enabled through the optical dipole potential: The force on an atom exposed to an off-resonant, spatially-varying intensity distribution is attractive (repulsive) when the laser is tuned below (above) the atomic resonance. This has led to numerous optical guiding schemes tailored for particular applications [1, 4-10]. Broadly speaking, red-detuned guides are simpler to create, but the high field confinement leads to higher photon scattering rates and level shifts; blue-detuned guides require beam shaping, but confine atoms to the low intensity regions of the beam and can significantly reduce photon scattering and other perturbations [11-15] necessary for sensitive measurements [16,17].

Atom guidance in HOFs has been demonstrated using both red- and blue-detuning, and each technique has benefits and disadvantages. Red-detuned guidance in a capillary has been done with both hot [4,5] and cold atoms [1,2,8], and is relatively easy to align. Recently, atoms have been guided in photonic crystal fibers (PCF) [1,2] in which the small mode-field area leads to strong atom-photon coupling and large optical depths. However, the high-intensity guides should be extinguished during the experiments to avoid energy level shifts during which atoms can escape. Blue-detuned, evanescent field guiding has been demonstrated in capillaries [6,7], but is inefficient since most of the guide laser power remains in the glass. Furthermore, because the evanescent field is at a submicron distance from the fiber wall, the field must be strong enough to overcome the attractive van der Waals force [6].

In this paper, we demonstrate atom guidance using our recent proposal [18] to use a higher order, blue-detuned hollow beam in a hollow waveguide, which both efficiently uses the guide light and provides a perturbation-reduced environment for the atoms. We compare guidance using the first three optical modes – the fundamental $HE_{11}$ mode, the azimuthally-polarized $TE_{01}$ mode, and the second order family of hollow modes. While we observe the highest atom number guidance using red-detuned light, we show that the blue-detuned beams guide atoms with a 10-fold reduction in the recoil scattering rate and that the blue-detuned guides can be loaded from a red-detuned beam inside the capillary for dark confinement with high atom flux, which may be useful for extending measurement time windows for tightly confined atoms [1].

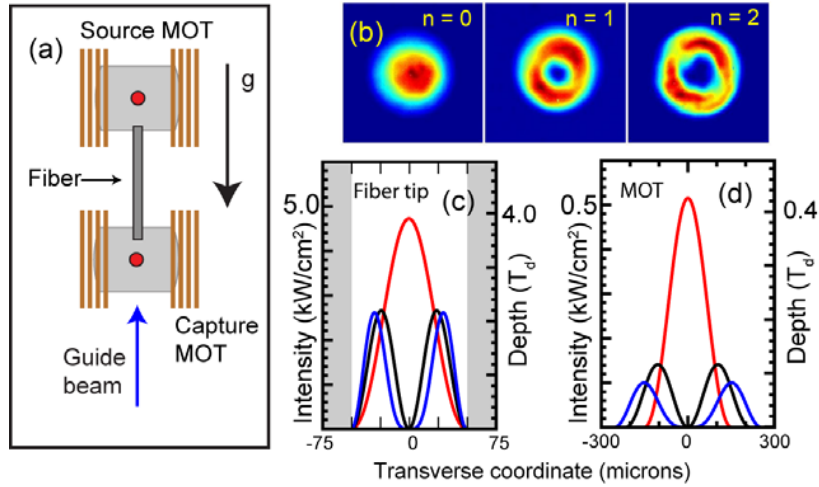

Fig. 1. (a) Experimental setup. (b) Beam profiles at the output of the hollow guide. Beam cross-sections at the guide output (c) and source MOT location (d). Plots also show the optical potential of the $n = 0$ (red), $n = 1$ (black) and $n = 2$ (blue) modes in units of the Doppler temperature $T_d = \hbar\Gamma/2$ for $\Delta = 1$ nm and 100 mW of input power. The gray shaded area in (c) represents the glass region of the capillary (core diameter = 100 μm).

## 2. Experimental setup

The experimental layout is shown in Fig. 1(a). A source magnetooptical trap (MOT) is situated 1.5 cm above the tip of a hollow, 3-cm-long, 100-micron-diameter hollow rod. Transported atoms are captured in a detection MOT $\approx$11 cm below the source MOT. The two MOTs are independently controlled, having separate anti-Helmholtz coils and laser beams, though because of their close proximity to one another, we use both continuous and pulsed bias coils as needed. The guide laser beam passes upward through the hollow rod. Atoms from the source MOT are loaded directly into this beam during the molasses stage, which cools the atom sample to ~10 μK.

In this work, we guide atoms through a hollow optical waveguide using the fundamental and first two higher order optical waveguide modes. The solutions to these modes are well known [19] and are only briefly described. The guide beams are derived from diode lasers operating within a few nanometers of the $^{85}$Rb $D_2$ line at 780.24 nm. The intensity distributions of the three modes considered in this paper are [19]:

$$I_n(r) = \frac{P_0}{\pi}\left[\frac{J_n\left(u_n\frac{r}{a}\right)}{aJ_{n+1}(u_n)}\right]^2 \qquad (1)$$

where $J_n(x)$ is the $n^{th}$ order Bessel function, $a$ is the radius of the capillary core, $r$ is the radial coordinate, and $P_0$ is the input power. For $n > 0$, these are hollow intensity distributions. The $u_n$ are the arguments producing the first finite zeros of the $n^{th}$ Bessel function. Throughout the paper, unless otherwise specified, we refer to the beams that produce these profiles by the value, $n$, of the subscript in Eq. 1. Experimental images at the tip of the guide output are shown in Fig. 1(b), and their cross sections are in Fig. 1(c). To lowest order, the radial intensity profile is quadratic for $n = 0$ and $n = 1$, and quartic for $n = 2$, scaling as $r^{2n}$ for $n > 0$. From the Virial Theorem, the time-averaged potential energy is $U_{avg} = K_{avg}/n$, where $K_{avg}$ is the time-averaged kinetic energy. For a given ensemble temperature, therefore, anharmonic profiles provide a reduced perturbation environment for low scattering rates [12, 15-17].

For a laser detuning larger than the hyperfine splitting, the optical potential can be described by [20]:

$$U(r) = \frac{\hbar \Gamma I(r)}{24 I_s}\left(\frac{\Gamma}{\Delta + \Delta_{LS}} + \frac{2\Gamma}{\Delta}\right) \quad (2)$$

where $\Delta$ is the detuning from the $D_2$ transition and $\Delta_{LS}$ is the fine structure splitting. $\Gamma/2\pi = 6.0$ MHz is the natural linewidth, and $I_S = 2.5$ mW/cm$^2$ is the saturation intensity for off-resonant, polarized light [21]. With 100 mW of power inside the guide, the peak intensities of $I_0(r)$, $I_1(r)$, and $I_2(r)$ are 4.7 kW/cm$^2$, 2.66 kW/cm$^2$, and 2.61 kW/cm$^2$. However, because the divergence increases for higher order modes, the beam diameters increase with $n$ at the MOT [Fig 1(d)], resulting in lower peak intensities and trap depths (0.52 kW/cm$^2$, 0.14 kW/cm$^2$, and 0.10 kW/cm$^2$ for $n = 0, 1, 2$, respectively). For $\Delta = 1.0$ nm and $P_0 = 100$ mW, the potential is high enough (0.08 $T_d$ for $n = 2$) to capture a significant thermal fraction of atoms from the MOT, which was cooled to ~10 μK ($\approx 0.07$ $T_d$) during the molasses stage. However, the MOT size is significantly larger than the beam diameters so only a small fraction of the MOT atoms are loaded into the beams. For near-resonant light, spontaneous Raman scattering is significant. When $\Delta \ll \Delta_{LS}$, an approximate form for this scattering rate is [20]:

$$\Gamma_{SP} = \frac{\Gamma}{12}\left(\frac{\Gamma}{\Delta}\right)^2 \frac{I_{AVG}}{I_s} \quad (3)$$

where $I_{AVG}$ is the time-averaged intensity sampled by the atoms. Blue-detuned traps can make $I_{AVG}$ small compared to the peak intensity depending on the trap potential form. In Ref. [14], the harmonic blue-detuned trap had a scattering rate reduced by 50 over a comparable red-detuned trap, and Ref. [15] used a box-like potential to achieve a reduction of 700.

The fundamental HE$_{11}$ mode in Eq. 1, $n = 0$, is formed simply by spatially filtering a laser beam with single mode optical fiber which is a close approximation to the HE$_{11}$ hollow fiber mode, but the two higher order modes require further beam shaping. To produce $I_1(r)$, we use the TE$_{01}$ cylindrical waveguide mode, which is azimuthally-polarized and is generated as described in Refs. [22,23]. Briefly, because this mode is closely matched to the first excited mode of a solid core optical fiber, we can use the output of a few mode optical fiber in which the TE$_{01}$ mode has been preferentially excited. This mode selection is done by passing a Gaussian beam through a vortex phase plate with a $2\pi$ azimuthal phase winding. When this modified beam is coupled into Corning HI-1060 fiber with cutoff wavelength of 980 nm, the fundamental HE$_{11}$ mode is eliminated by the phase winding of the input beam, and the correct cylindrical vector beam can be selected by adjusting the polarization with an inline polarization controller [22].

The next higher order family of modes with $n = 2$ ($HE_{31}$, $EH_{11}$, and $HE_{12}$) has an intensity profile proportional to $I_2(r)$ in Eq. 3. The exact profiles could, in principle, be generated in a similar manner to the $TE_{01}$ mode using a different solid core optical fiber with larger cutoff wavelength, but mode selection from the solid fiber becomes more difficult as the core size grows. Alternatively, one could use subwavelength grating structures [24] if exact polarization profiles are needed. In this work, we generate a beam with the approximate intensity profile using a phase plate with a $4\pi$ phase winding. While this has a spatially uniform polarization profile, the overlap integral with this family of modes is calculated to be 68%. The remaining 32% overlap occurs with other higher order hollow modes of the guide. For coaxial alignment into the rod, modes that are not hollow are only excited through aberrations of the input beam.

The large core size of the hollow waveguide demands proper mode matching to eliminate speckle and excitation of other modes, so the incident beam size of the three input beams is carefully adjusted to have the correct size at the hollow rod tip. We use a Pentax C60812 8-48mm zoom lens to collimate the output of the light delivery fibers. The beams are focused by a 200mm achromat, mounted outside the chamber, into the bottom of the hollow rod, and to account for slight variations in focal position for the three beams, the axial lens position is adjusted by a micrometer. The fundamental beam has the highest optical transmission of ~80%, while for $n = 2$ we obtain a ~45% throughput. The attenuation lengths for $I_n(r)$ are $\alpha_0 = 48.2$ cm; $\alpha_1 = 30.9$ cm; and $\alpha_2 = 10.6$ cm, giving calculated transmissions of 0.94, 0.91, and 0.75, respectively. Our transmission of $I_2(r)$ is significantly less than 0.75, most likely due to the overlap with higher order modes that will have even shorter attenuation lengths, and to larger coupling losses at the input, which are expected because the beam diameter is larger. Significantly longer attenuation lengths can be achieved by metal-coating the interior wall of the optical guide [18]. We note that the ends of the rod are polished and coated with aluminum so that light that has escaped from the core cannot interfere with the loading process.

## 3. Experiments

The guiding beams are kept on during the MOT loading and molasses stages. We consider time $t = 0$ to be the end of the molasses stage when the atoms first begin freefall into the guide. Our signal is the fluorescence from atoms captured in a second MOT below the capillary, and is proportional to the guided atom number.

*3.1 Guiding as a function of detuning*

In this section, we compare atom guiding through the hollow rod with red- and blue-detuned light using the three intensity profiles of Eq. 1. For $n = 1$ and $n = 2$, we have used blue-detuned guiding, which should show greatly reduced scattering compared to the red-detuned guiding of $n = 0$ when the trap depth is high. The qualitative differences between the guides and the effects of scattering are shown in Fig. 2, which plots the guided atom number as a function of $\Delta$. We show two different guide powers for each of the three modes.

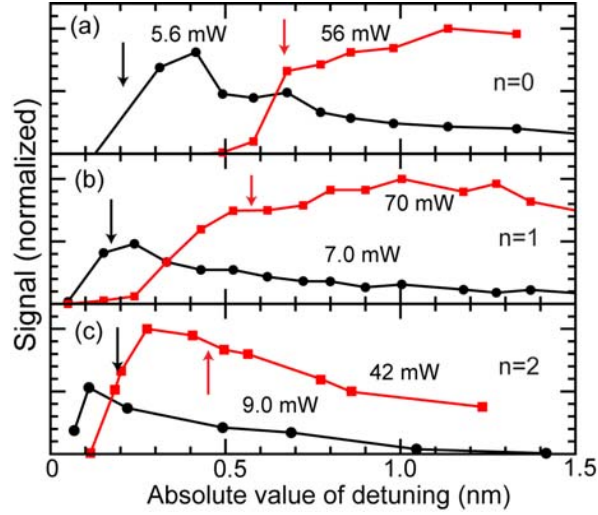

Fig. 2. Comparison of atom flux for the different beam types. (a) Red-detuned $HE_{11}$ mode; (b) blue-detuned $TE_{01}$ mode; (c) blue-detuned $n = 2$ mode. We have used two different guide powers for each case as indicated. Arrows indicate $\Delta_g$, the detuning at which the peak scattering force equals the force of gravity. Signals are normalized for each beam type independently; relative atom numbers are discussed in the text. Note that the detuning values for $n = 0$ (red-detuning) are negative.

The figure shows normalized atom flux for the $n = 0$, $n = 1$, and $n = 2$ beam profiles. For each case, the effects of photon scattering are apparent, but the magnitude of the effect varies. In this study, the relevant parameter is the intensity of the beam inside the capillary core, so we have kept this parameter approximately the same for the three beam shapes. This does, however, lead to significantly different beam intensities at the MOT [see Fig. 1(d)], leading to variations in atom flux between the beam types, so they are plotted separately.

Our guide laser beam propagates upward through the detection chamber into the source chamber. Thus, for sufficiently high scattering rates that occur at very small $\Delta$, the atoms cannot propagate through the capillary, either because they scatter enough photons to boil out of the guide beam potential, or because photon pressure from off-resonant scattering overcomes gravity. This pressure can be useful for manipulating atom velocities inside optical guides: Atom levitation due to guide radiation pressure was observed at very small detunings in Ref. [13], and additional near-resonant beams have been suggested for controlling atom motion inside PCF [2]. For a particular mode, the atom flux is determined by the depth of the optical potential at the source MOT and photon scattering effects. Without photon scattering, one would see increased atom flux for smaller $\Delta$ because the trap depth would increase; however, as shown the atom flux is reduced due to increased photon scattering.

The effects of the scattering force are clearly observed using red-detuning, shown in Fig. 2(a). At high power (56 mW), no atoms are guided through the capillary until $|\Delta| \approx 0.6$ nm. We define $\Delta_g$ as the detuning at which the peak scattering force equals gravity, indicated by the arrows in Fig. 2. For red-detuned guiding, atom transport is observed when $|\Delta| = \Delta_g$. When 5.6 mW of guiding power is used (black curve), the atoms are again guided when $|\Delta| = \Delta_g$, which is reduced to 0.2 nm at this power. Since atom transport begins at $|\Delta| = \Delta_g$ for red-detuning, it is clear that the atoms are primarily in the regions of peak intensity. For blue-detuned guiding in Figs. 2(b) and 2(c), however, atom transport occurs for $|\Delta| < \Delta_g$, with the effect for $n = 2$ being more pronounced. This shows that the time-averaged intensity sampled by the atoms is lower. Quantitative measurements of the scattering rates are shown in section 3.2.

The combined effects of trap depth and scattering lead to an intensity-dependent maximum of the atomic flux. We note that because the guiding beam is directed upward, the detuning at which atoms begin to be guided is larger than if it were directed downward because in the latter case the scattering force would be in the same direction as gravity. Reduced atomic flux can also occur due to heating atoms over the potential barrier, but as discussed in Section 3.2, the heating for $\Delta > \Delta_g$ is insignificant over our 3 cm guide length.

For similar intensities at the MOT, the red-detuned $n = 0$ mode guides $\sim 10^6$ atoms through the capillary, approximately 5x more atoms than the blue-detuned $n = 2$ mode, and $\sim 10$-15x more atoms than the $n = 1$ mode. There are two main reasons for this difference. First, for a blue-detuned guide, the relevant parameter is the *minimum* peak intensity of the guide boundary – aberrations in the beam may lead to "leaky" pathways for the atoms to exit. For a red-detuned beam, however, the atoms can remain bound in the high intensity portions of the beam. Second, loading into optical traps is generally more favorable for red-detuned guides, and experiments with optical traps have observed that atoms are loaded with higher density if the trap is red-detuned [11, 25] but these density enhancements are not observed with blue-detuned light. As we suggested in our original proposal [18], if guided atom flux is most important, it is best to use red-detuned guidance, but as we show in the next section more quantitatively, the blue-detuned guidance offers much better reduction of photon scattering. The blue-detuned guides can also be loaded from the red-detuned $HE_{11}$ mode, as shown in Sec. 3.3, so that increased atom flux and low scattering are achieved.

*3.2 Photon scattering rates*

Photon scattering rates in optical potentials are often determined experimentally through state-selective detection: Atoms are first optically pumped into the lower hyperfine manifold, and their relaxation rate into the upper hyperfine manifold is measured [14, 15, 26]. Although we could perform a similar spectroscopic measurement within the waveguides, the unidirectional optical guide makes it straightforward to determine the effective force on the atoms by simply measuring the time dependence of the atom flux into the detection chamber, because the upward radiation pressure slows the atoms. This technique has previously been used for small detunings with high scattering rates [13, 27] and gives an accurate measurement of the recoil photon scattering rate, whereas the spectroscopic measurement only measures the spontaneous Raman scattering rate. The recoil scattering rate scales with $1/\Delta^2$; for $\Delta < \Delta_{LS}$, the spontaneous Raman scattering rate also scales with $1/\Delta^2$, but for $\Delta > \Delta_{LS}$ scales with $1/\Delta^4$ [26].

Typical time-domain curves of guided atom signal through the capillary are shown in Fig. 3. Here, we extinguish the laser guide with variable delay from $t = 0$ to $t = 200$ ms; atoms that exit the hollow rod prior to the shutoff are captured in the collection MOT, while those remaining in the guide are lost by hitting the glass walls. Since the end of our guide is 46 mm below the source MOT, the atoms with zero downward velocity at $t = 0$ will exit the capillary at $t \approx 97$ ms; near this shutoff time, the integrated atom flux increases most quickly. The curves in Fig. 3 depend strongly on $\Delta$. In particular, as $\Delta$ decreases, the increased scattering force causes atoms to take longer to fall through the guide into the detection chamber. The shape of this integrated atom signal depends on the initial MOT distribution and the temperature, but to calculate the scattering force, we have assumed a point source of atoms with a Maxwell-Boltzmann velocity distribution along the capillary axis:

$$P(v)dv = C \exp\left(\frac{-Mv^2}{2kT}\right)dv \qquad (4)$$

Because the position at a later time is simply $y = y_0 + vt - 0.5gt^2$, we can solve for $v$ to find the atom flux through the end of the capillary as a function of $t$ and fit the results to our data. We note, however, that any approximate functional form, applied consistently to these data, results in similar arrival times. The integrated flux of the velocity distribution fits well to the following model:

$$N(t) = A + B\,\mathrm{erf}\left(\frac{t - t_0}{\tau}\right) \tag{5}$$

where $A$ and $B$ are constants, $t_0$ is the travel time, and $\tau$ is the characteristic width of the integrated flux curve.

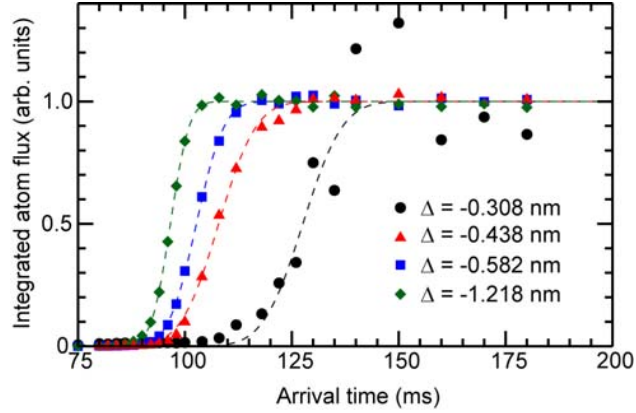

Fig. 3. Integrated atom flux through the end of the capillary guide as a function of time. Plots are shown for red-detuned, $n = 0$ guidance at different detunings. Dashed lines are fits using an error function described in the text. For clarity, error bars are not shown, but are approximately the same size as the symbols for $\Delta = -1.2$ nm, $-0.58$ nm, and $-0.44$ nm. For $\Delta = -0.31$, reduced atom flux gave error bars ~20%.

Extracting the arrival times, $t_0$, for each of the beam types as a function of $\Delta$, we can determine the average acceleration of the atoms. The difference between this acceleration and gravity, $g$, is the deceleration caused by photon scattering, $\gamma_{sc}v_r$, where $v_r = 5.88$ mm/s is the recoil velocity of $^{85}$Rb and $\gamma_{sc}$ is the scattering rate. In Fig. 4, we have plotted $\gamma_{sc}$ for the different beam types. The curves follow the expected $1/\Delta^2$ proportionality shown in Eq. 3, as indicated by the dotted line fits. If we write $I_{AVG}$ from Eq. 3 as $I_{AVG} = \beta I_{max}$, where $I_{max}$ is the peak intensity of the beam, we find the relative reduction of the average intensity, $\beta$, of the red-detuned $n = 0$, and blue-detuned $n = 1$ and $n = 2$ modes to be 0.44, 0.064, and 0.041, respectively. Thus, the blue-detuned $n = 2$ mode has >10x lower scattering rate than the red-detuned $n = 0$ mode for our guide parameters. We also show the curve for $\beta = 1.0$, which is the scattering rate at peak intensity (black solid line).

Measuring scattering rates at small detunings through the recoil force is not new [13, 27], but we note that this technique appears to be quite effective at detecting low scattering rates at much larger detunings as well – the atoms are only in the capillary for less than 100 ms, so at large detunings with scattering rates near 100 s$^{-1}$, only ≈10 scattering events occur. We note that this measurement assumes a constant scattering force throughout the capillary, which of course is not valid for very low scattering rates when only a few photons are scattered during transit. While the atoms are in the guide, they are heated by approximately $T_r\gamma_{sc}$, where $T_r = 350$ nK is the recoil temperature increase on each scattering event. Because the time inside the guide is only ~0.1s, the atoms are heated by only a small fraction of the potential depth (several hundred microKelvin) even at small detunings.

It should be possible to use higher $n$ values to guide atoms with lower scattering rate [12, 20], although the attenuation length may become too short unless the capillary is coated [18]. In our case with a vertical capillary, the optical potential does not support the atoms against gravity, so we would expect higher scattering rates if the capillary is oriented horizontally. A horizontal orientation might be desirable, however, for increased interrogation time.

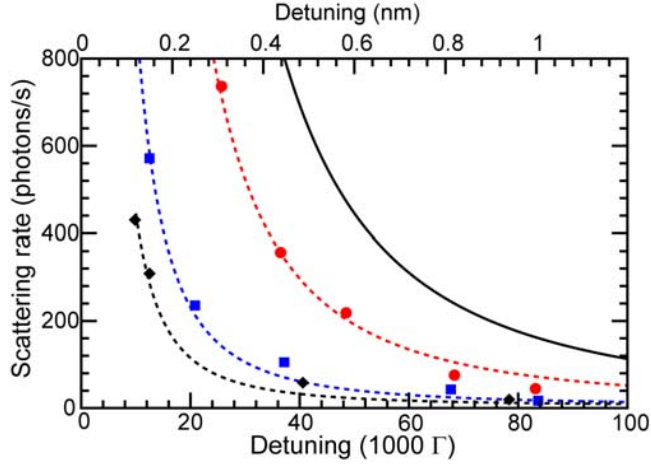

Fig. 4. Scattering rate versus detuning for the fundamental red-detuned beam (red circles), $TE_{01}$ blue-detuned mode (blue squares), and second excited mode (black diamonds). Fits are shown as dashed lines. The solid black curve is the calculated scattering rate at the peak intensity.

*3.3 Red-detuned loading of the blue-detuned guide.*

For highest atom flux, red-detuned guiding is most efficient. Unfortunately, to perform experiments on atoms in a perturbation-reduced environment, one must extinguish the guide light during the measurement time. This allows only a brief window during which the measurement can be performed before atoms are lost [1] and will also lead to measurement-time broadening of spectral features. To increase the measurement time, we consider transferring atoms from the red-detuned $n = 0$ guide into a copropagating $n = 2$ blue-detuned guide to provide increased optical densities, a reduced perturbation environment, and longer measurement times.

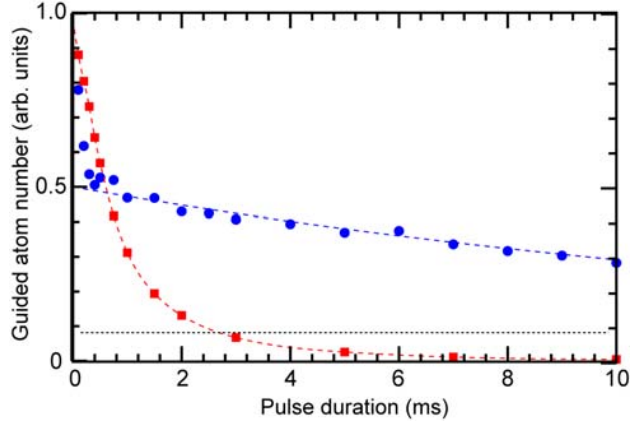

Fig. 5. Guided atom number as a function of extinction time, $T_{off}$, of the red-detuned guide. When the blue-detuned beam is not present (red points) the atoms quickly escape to the capillary walls. When the blue-detuned beam is present to support the atoms (blue points), the atoms can be later recaptured by the red-detuned guide even for long extinction times.

In Fig. 5, we measure the confinement time with and without the blue-detuned beam by a release-and-recapture method: At the time when the atoms enter the capillary ($t \approx 60$ ms), the red-detuned light is extinguished for a brief period, $T_{off}$, and then turned back on, and recaptured atoms make it through the capillary and are detected as before. Without any blue-detuned guide present, this signal decays with a $1/e$ time constant of 1.5 ms due to atoms that strike the capillary wall. If we turn the blue-detuned beam on when the red guide is shut off, the atoms cannot escape and are recaptured by the red-detuned beam. For $T_{off} < 1$ ms, there is an initial large loss of signal due to the size of the blue-detuned beam: Any atoms outside the peak-peak diameter of ~60 μm are lost (see Fig. 1c). After this initial loss, the remaining atoms are confined in the blue-detuned guide with gradual loss until they exit the capillary. We note that for these experiments the relative guided atom number for the $n = 2$ beam in the absence of the red-detuned $n = 0$ beam was near 0.1, as indicated by the dashed black line in Fig. 5, so the short time enhancement ($T_{off} < 5$ ms) due to the blue-detuned beam is about 5x. We note that because hollow beams have been successfully propagated through PCF [28, 29], it might be possible to use them to extend measurement times on atoms confined to PCF.

During $T_{off}$, the atoms continue falling in the blue-detuned hollow mode and sample more of the hollow beam. If the mode quality had deteriorated and developed potential minima along the capillary length, we would have expected the guided atom number to drop at the $T_{off}$ values corresponding to these locations. We did not observe this, and because the output mode quality was also good, the mode quality was likely good throughout the capillary.

## 4. Conclusions

We have guided cold atoms using the first three optical modes of a 100-micron-diameter capillary over a distance of 3 cm. Specifically, using time-of-flight measurements, we have observed a 10x reduction in photon scattering using the second excited, blue-detuned hollow mode compared with red-detuned guiding in the fundamental mode. We have also shown that red-detuned loading of a blue-detuned hollow mode can be useful for improved measurement time in perturbation reduced environments with increased atom flux. These results should be of interest for low power nonlinear optics, especially when extended to PCF confinement. This work was supported by the Office of Naval Research and the Defense Advanced Research Projects Agency. We gratefully acknowledge helpful discussions with Guy Beadie and technical support from Barb Wright and Gary Kushto.